\RequirePackage{fix-cm}
\documentclass[smallextended]{svjour3}       
\smartqed  
\usepackage{graphicx}
\begin{document}

\title{125 - 211 GHz Low Noise MMIC Amplifier Design for Radio Astronomy
}


\author{Daniel White \textsuperscript{1,2}         \and
        William McGenn \textsuperscript{1,3} \and Danielle George \textsuperscript{1,2} \and Gary A. Fuller \textsuperscript{1,3}
        \and Kieran Cleary \textsuperscript{4} \and Anthony Readhead \textsuperscript{4}
        \and Richard Lai \textsuperscript{5} \and Gerry Mei \textsuperscript{5}
}


\institute{Daniel White \at
              \email{daniel.white-4@manchester.ac.uk} 
              \and
\at \textsuperscript{1} Advanced Radio Instrumentation Group (ARIG)
\at \textsuperscript{2} School of Electrical and Electronic Engineering, University of Manchester, Manchester, UK
\at \textsuperscript{3} Jodrell Bank Centre for Astrophysics, School of Physics and Astronomy, University of Manchester, Manchester, UK
\at \textsuperscript{4} Department of Astronomy, California Institute of Technology, Pasadena, CA, USA
\at \textsuperscript{5} Northrop Grumman Aerospace Systems, Redondo Beach, CA, USA
}

\date{Received: date / Accepted: date}

\maketitle

\begin{abstract}
To achieve the low noise and wide bandwidth required for millimeter wavelength astronomy applications, superconductor-insulator-superconductor (SIS) mixer based receiver systems have typically been used. This paper investigates the performance of high electron mobility transistor (HEMT) based low noise amplifiers (LNAs) as an alternative approach for systems operating in the 125 --- 211 GHz frequency range. A four-stage, common-source, unconditionally stable monolithic microwave integrated circuit (MMIC) design is presented using the state-of-the-art 35 nm indium phosphide HEMT process from Northrop Grumman Corporation. The simulated MMIC achieves noise temperature (T\textsubscript{e}) lower than 58 K across the operational bandwidth, with average T\textsubscript{e} of 38.8 K (corresponding to less than 5 times the quantum limit (hf/k) at 170 GHz) and forward transmission of 20.5 $\pm$ 0.85 dB. Input and output reflection coefficients are better than -6 and -12 dB, respectively, across the desired bandwidth. To the authors knowledge, no LNA currently operates across the entirety of this frequency range. Successful fabrication and implementation of this LNA would challenge the dominance SIS mixers have on sub-THz receivers.



\end{abstract}

\section{Introduction}
\label{intro}


High electron mobility transistors (HEMTs) have been important components in many areas of electronics, replacing gallium arsenide (GaAs) metal-semiconductor field effect transistors in low noise amplifiers (LNAs) \cite{mimura}. Continued HEMT development in terms of device composition, semiconductor substrates, and transistor features (such as reduced gate length \cite{palacios} and T-shaped gates \cite{lee}) have improved the high frequency performance of HEMTs. These properties make HEMT based LNAs a viable technology for front-end amplification in millimetre wave receivers for radio astronomy.

Historically, superconductor-insulator-superconductor (SIS) mixers have been favoured at millimetre and sub-millimetre frequencies for their ability to provide low noise measurements across a wide bandwidth. For example, SIS mixers configured as heterodyne receivers are utilised extensively and successfully in the Atacama Large Millimeter/submillimeter Array (ALMA) between 84 GHz --- 1 THz \cite{alma3,alma10}. However, SIS mixers can only operate when cryogenically cooled to 4 K which involves complex cooling systems with high operational costs \cite{cuadrado2}.

Advances in HEMT technology have produced that LNAs are able to operate across similar broad bandwidths to SIS mixers above 50 GHz. Scaling down transistor feature dimensions, precise fabrication and using high electron mobility semiconductor materials results in better transistor performance not only in high frequency gain, but also low noise characteristics \cite{suemitsu}. Whilst originally developed with gallium arsenide (GaAs) substrates, HEMTs fabricated on indium phosphide (InP) substrates have shown even greater electron mobility. In addition to this, when cryogenically cooled to 15 K, InP based transistors have been shown to exhibit noise performances just 4 --- 5 times the quantum limit \cite{byerton}. 

These advances have prompted the investigation of using LNA's at higher frequencies, extending into the frequency ranges currently covered by SIS mixers. One such example is the W-band LNA designed for operation across 67 --- 116 GHz \cite{cuadrado} which achieved noise temperature (T\textsubscript{e}) of lower than 28 K when cryogenically cooled to 15 K. When compared to the state-of-the-art SIS mixers used in ALMA, the LNA operated with lower noise across its bandwidth.


In this paper, we investigate the viability of LNAs operating in the 125 --- 211 GHz frequency range. Table \ref{tab:1} shows the target LNA specification proposed by the European Southern Observatory. The specification is based on exploring the performance limits of semiconductor devices operating within the frequency ranges of 125 --- 163 and 163 --- 211 GHz, and has been drawn from existing specifications in order to be comparable to current state-of-the-art SIS-based receivers \cite{asayama,belitsky}. The noise temperature requirement is specified as the maximum value allowed across 80 \% of the bandwidth (an aggregate of 68.8 GHz of the band between 125 GHz and 211 GHz) and the maximum allowed at any frequency in the band. Since this frequency range is currently  split between two separate receivers on ALMA, the noise requirements are specified for these two contiguous bands separately. 

A monolithic microwave integrated circuit (MMIC) design is presented based on the 35 nm gate length InP HEMT process introduced in 2007 by Northrop Grumman Corporation (NGC) \cite{35nm}. At the time of its inception, this technology produced a single transistor stage amplifier that provided amplification up to 300 GHz. It has since been used to produce amplifiers with state of the art noise performance operating in the sub-THz range when cryogenically cooled \cite{lna1,lna2,lna3,lna4}. This process represents the best recorded cryogenic sub-THz noise performance for any currently available transistor technology.

\begin{table}
\caption{Low Noise Amplifier Performance Specification}
\label{tab:1}       
\begin{tabular}{l l l l}
\hline\noalign{\smallskip}
Frequency Range & Noise Temperature & Gain & S11 \& S22\\
\noalign{\smallskip}\hline\noalign{\smallskip}
125 - 163 GHz & $<$ 40 K across 80 \% Band & 35 - 40 dB & $<$ -10 dB \\
 & $<$ 60 K across 100 \% Band & Less than 6 dB ripple & \\
163 - 211 GHz & $<$ 45 K across 80 \% Band & Peak-to-peak &\\ 
& $<$ 75 K across 100 \% Band & &\\ 
\noalign{\smallskip}\hline
\end{tabular}
\end{table}

\section{Low Noise Amplifier Design}
\label{sec:1}

\begin{figure}
  \includegraphics[scale=0.32]{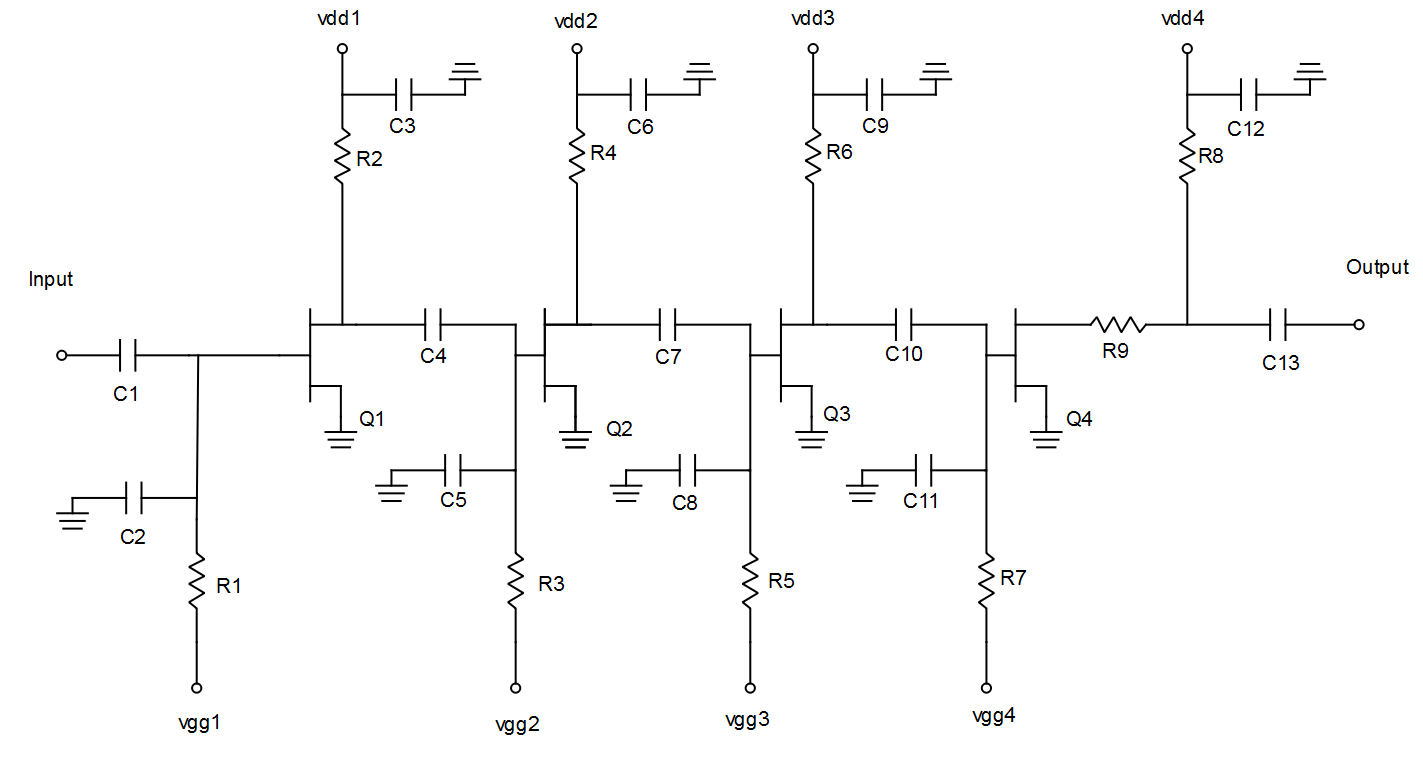}
\caption{Simplified Circuit Schematic}
\label{fig:3}       
\end{figure}

In this section we discuss the design of the MMIC. The presented results are simulated at a physical temperature of 20 K using Keysight's Advance Design System (ADS) with Momentum electromagnetic simulations of passive matching networks \cite{ads}. The amplifier consists of four transistor stages arranged in common-source topology. All the transistors are two-finger devices with gate width of 10 $\mu$m, chosen due to their low noise performance across the operational bandwidth. DC bias lines and matching networks are designed using microstrip transmission lines, resistors and capacitors. Series blocking capacitors are used at the input and output of the amplifier to prevent DC bias signals from traversing the circuit interfaces. A series resistor in the output matching network provides circuit stabilisation. A simplified circuit schematic is shown in figure \ref{fig:3}.

The first transistor stage is optimised to achieve low noise across the operational bandwidth. The gain of the first stage however must be sufficient enough for the subsequent stages to have a negligible effect on the overall T\textsubscript{e}, as illustrated in the Friis cascaded system equation \cite{friis}. As the gain of a single stage is typically less than 10 dB at such high frequencies, the second stage is also optimised to achieve low noise. The third and fourth stages have less impact on the overall noise performance, and so are optimised for achieving flat gain and to reduce the output reflection across the operational bandwidth. Each transistor has independent gate and drain bias lines, allowing each stage to be tuned in order to achieve optimum noise and gain performance.

\section{Simulation Results}
\label{sec:3}
\begin{figure}
  \includegraphics[width=1\textwidth]{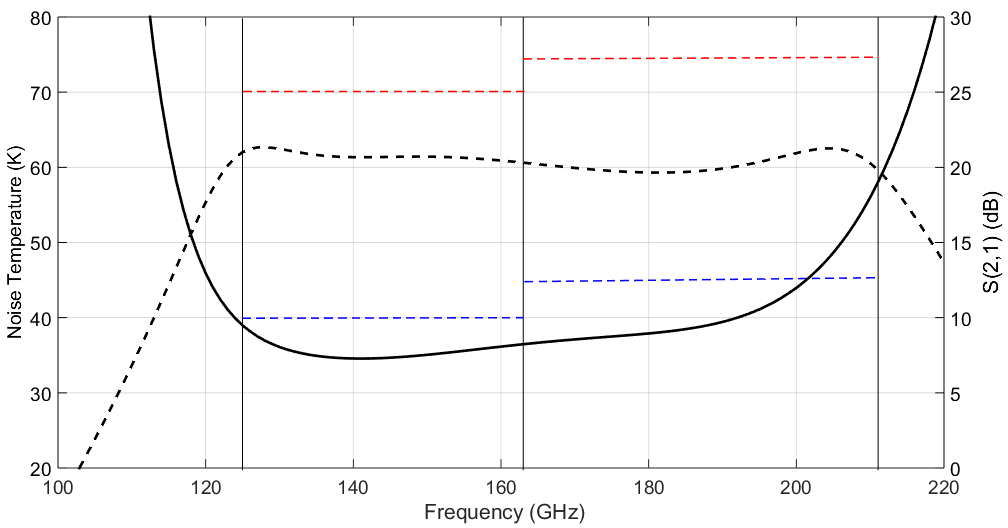}
\caption{Noise Temperature (solid) and Forward Transmission (dotted) simulated results. Dotted blue and red lines indicate the 80 and 100 \% noise temperature LNA specifications, respectively, for each sub-band.}
\label{fig:1}       
\end{figure}

Figure \ref{fig:1} shows the T\textsubscript{e} (solid line) and forward transmission (S\textsubscript{21}) (dotted line) plotted against frequency. The amplifier achieves T\textsubscript{e} lower than 58 K across the entire bandwidth, with an average T\textsubscript{e} of 38.8 K, corresponding to less than 5 times the quantum limit (hf/k) at 170 GHz. The S\textsubscript{21} of the amplifier is 20.5 $\pm$ 0.85 dB.

The simulated T\textsubscript{e} across 125 - 211 GHz surpass the requirements set out in the specification summarised in table \ref{tab:1}. Across 125 --- 163 GHz the T\textsubscript{e} is under 40 K with minimum T\textsubscript{e} of 34.6 K. Across 163 --- 211 GHz the T\textsubscript{e} is under 45 K for greater than 80 \% of the bandwidth, with a maximum T\textsubscript{e} of 57.9 K. These simulations indicate that the fabricated MMIC would be capable of meeting the noise performance requirements.

The simulated S\textsubscript{21} of the amplifier averages 20 dB across the desired bandwidth, falling short of the 35 - 40 dB specification shown in table \ref{tab:1}. In order to increase the gain to the specification, two LNA modules can be connected together though a microwave isolator. The isolator prevents signal reflections between modules, which would cause excessive ripple in the S\textsubscript{21} measurements.

\begin{figure*}
\includegraphics[width=1\textwidth]{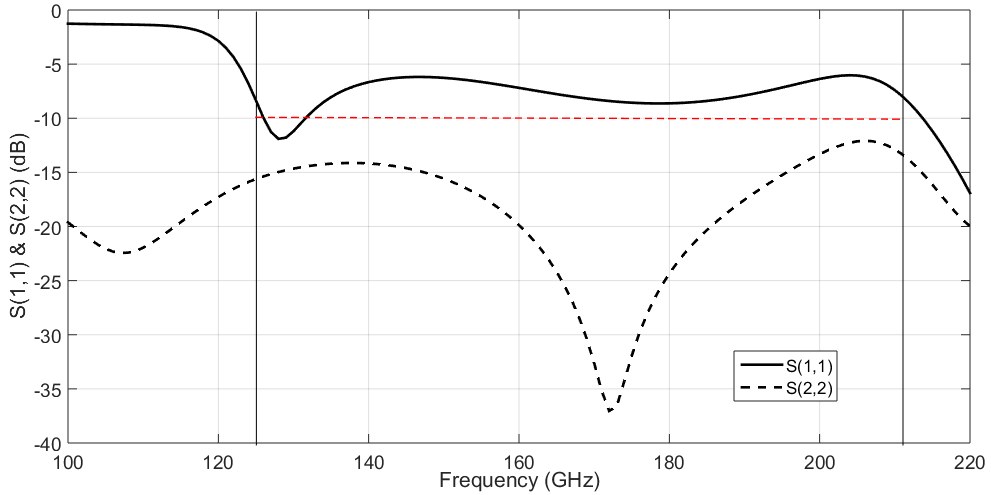}
\caption{Input (solid) and Output (dashed) Reflection Coefficients. Dotted red line indicates the reflection coefficient LNA specification}
\label{fig:2}       
\end{figure*}

The input (solid) and output (dotted) reflection coefficients are shown in figure \ref{fig:2}. The input and output reflection coefficients are better  than -6 and -12 dB across the combined bandwidth of 125 --- 211 GHz. Reverse isolation (S\textsubscript{12}) reaches a maximum of -47 dB at 211 GHz, demonstrating a sufficient level of isolation between the output and input of the amplifier. Simulating the LNA over a frequency range from 0 to 400 GHz shows that the stability factor is greater than 7 at all frequencies, indicating that the amplifier is unconditionally stable.

Previous results from this process indicate that the measured performance matches the simulated performance closely \cite{cuadrado}, demonstrating that the 35 nm InP HEMT process is capable of achieving the desired LNA performance. This MMIC will be included on a future wafer run, and will be cryogenically tested to verify simulation results. Successful fabrication of the MMIC will indicate that LNAs are capable of operating as front-end receivers in the millimetre-wave spectrum.

\section{Discussion}
\label{sec:4}
The noise performance of the presented MMIC is comparable to two SIS mixers operating within this bandwidth \cite{asayama,belitsky}. This, combined with a uniform gain performance across the bandwidth, suggests that an LNA could find application in ALMA to combine the operating bands of 4 + 5 (125 --- 211 GHz) into a single ultra wideband receiver. This would help to reduce the operational costs associated with running the observatory, as it would reduce two 4 K coolers to a single 15 K cooler. It would also make an additional slot available on the front-end cryogenic system due to the combination of two existing slots.

Additional components in the ALMA receiver cartridge such as the feed, orthomode transducer (OMT) and the coupling optics (whether warm as for ALMA band 4 \cite{asayama} or cold as for band 5 \cite{belitsky}) all have a contribution to the total receiver noise, and are included in the ALMA receiver noise budget. A direct and fair comparison of an LNA- and a SIS-based system will require an LNA assembled within an ALMA cryostat, which will include feedhorn, OMT and optics noise contributions. However, the results presented in this paper indicate that there is considerable potential for an LNA based receiver to have competitive performance with a SIS-based receiver.

\section{Conclusion}
\label{sec:5}

This paper has presented a MMIC design capable of operating across the bandwidth of 125 --- 211 GHz, satisfying the LNA specification set out in table \ref{tab:1}. Across the  bandwidth of 125 --- 211 GHz the MMIC achieves T\textsubscript{e} less than 58 K, with maximum and minimum T\textsubscript{e}'s of 57.9 and 34.6 K, respectively. The average S\textsubscript{21} across the bandwidth is 20.5 $\pm$ 0.85 dB. Two LNA modules can be connected in series with a microwave isolator in order to increase the gain to that specified in table \ref{tab:1}. Input and output reflection coefficients are better than -6 and -12 dB, respectively, and S{\textsubscript{12} is better than -47 dB. The Rollett stability factor is greater than 7 at all frequencies indicating unconditional stability. The circuit will be included in a future 35 nm InP process wafer run to verify the simulations with measured results.

Following the fabrication process, the functioning MMICs will be tested on-wafer and in-module for S-parameters, noise figure, and linearity tests including 1 dB compression point, third order intercept and power dissipation. For in-module testing, an LNA body will be designed and populated with the ancillary components necessary for operation in a receiver. In particular, a waveguide-to-microstrip transition interface must be carefully designed in order to maximise the transmission of the input signal. Successful fabrication and measurement of the MMIC presented in this paper would challenge the dominance SIS mixers hold on millimeter-wave receivers for radio astronomy up to at least 211 GHz. Upgrading ALMA by replacing the two current 4 K SIS-based receiver cartridges required to cover the 125 --- 211 GHz frequency range with a single LNA-based cartridge operating at 15 K, and providing a broader instantaneous bandwidth would reduce the operational cost of the receiver, as well as potentially offering other operational and scientific advantages.

\begin{acknowledgements}
This study is funded by the Engineering and Physical Sciences Research Council Doctoral Training Partnership and the European Southern Observatory Technical Development Fund contract ESO/17/11279/ASP. The authors would like to acknowledge the contributions of Northrup Grumann Corporation and Caltech. We would also like thank Dr Gie Han Tan for his support of the work presented here.
\end{acknowledgements}



\end{document}